# Primordial particles and waves in the early Universe

Xiang Liu 1,2

1. Xinjiang Astronomical Observatory, Chinese Academy of Sciences, 150 Science 1-Street, Urumqi 830011, China
2. Physics and Astronomy Department, Qiannan Normal University for Nationalities, Duyun 558000, China

## Abstract

The observational evidence points to the origin of our Universe from a big-bang explosion, the normal matter we observed can be well explained by the particles created in the strong and weak interacting phases of the early universe. The dark energy and dark matter, which occupy major contents of the universe, may be created from earlier times than that of quarks. We propose that the dark energy is an integration of the gravitational waves (GWs) in early universe, and the dark matter is composed of x particles which are created earlier and have much heavier mass than quarks in early universe. The x particles may be bosons with no or little interaction with normal matter except gravity, and they can be free particles or Bose-Einstein Condensed (BEC) matter as the dominant matter in the galaxy clusters and galactic halos. This paper notes as a framework of ideas need to be investigated further.

1. Introduction

In the standard model of particle physics and symmetry, there are fermions, bosons, particles and anti-particles. They are quarks (u, d, c, s, t, b), leptons (e, μ, τ, $v_e$, $v_\mu$, $v_\tau$), bosons(γ, Z, W, gluon, Higgs). If these elemental particles come from early universe, one should study the primordial particles created in the early universe.

It is widely believed that our universe comes from a big-bang inflation, the atomic matter e.g. hydrogen is formed very after the first 3 minutes of the big-bang (Weinberg, 1972). Before that time, the universe is a kind of soup of the fermions and bosons and they are coupled because of high temperature. In much earlier time than the 3 minutes from the big-bang, the universe could be the plasma of elemental fermions and bosons which are fully coupled, and the universe is dark, only gravitational waves may be able to partly escape from it.

2. X particles in early universe

   (1) x particles and dark energy

The particles can be produced by pair creation through vacuum fluctuations in the early time of the big-bang inflation, they should have equal quantity of particles and anti-particles. When the temperature decreases, the weak-interaction will lead to the CP violation, so that the matter (consisting of normal particles) becomes slightly more than anti-matter (consisting of anti-particles), and so the remnant (normal matter) is left in our universe.

For the higher vacuum energy in very early universe, it is possible that the higher energy particles are created earlier. In this idea, the strong interaction particles are created earlier than weak-interaction particles, e.g., the quarks are created earlier than electrons. However, the first generation of particles might not be the quarks, we assume it is a particle x. The x particle should be heavier than quarks, the anti-x particle is the x particle itself; and when the temperature decreases, they will not be fully annihilated, but remain until today. The so-called dark matter could be consisting of the x particles.

Because x particles are created earlier than quarks and leptons, so x particles would not play an important role in the strong-interaction and weak-interaction. Therefore, with the higher energy or mass of x particles, they will be the best candidate of dark matter, which occupies about 23% of the energy of universe. The normal matter which created after the x particles, have lower energy or mass, occupies about 4% of the energy of universe. In the standard cosmology model, the so-called dark energy occupies about 73% of the energy of universe. For the dark matter is only weakly interacting with 'normal matter', the x particles may be bosons with no charges e.g. electron charges.

The dark energy occupies the majority of energy of the universe, it could be mostly created earlier than the x particles. It can be the vacuum fluctuation itself and releases its energy as gravitational waves. The dark energy cannot be electromagnetic waves and neutrinos which are formed after the x particles.

The primordial gravitational waves (PGWs) are not observed yet, but the gravitational waves from the BH-BH mergers have been detected by the LIGO/Virgo instruments, which proved the theory of gravitational waves. The GWs produced before x particles, may have higher energy than that from the BH mergers, but it is not clear if the frequency of the primordial GWs is higher than the stellar BH mergers or not, which depending on the fluctuation scales of space-time in early universe. The primordial GWs should have a background GWs distributed everywhere in the universe, but it may be difficult to be detected if there are astrophysical foreground GWs. It is possible that the primary GWs have a spectrum from very high to low frequencies, a power-law GW spectrum of energy distribution (SED) is assumed, $E = E_0 t^{-n}$ for the inflation of early universe. Here we assume the GW energy is anti-correlated with the scale of the inflating universe, i.e. the frequency of GW (and therefore its energy) is inversely proportional to the scale (or time) of baby universe. The dark energy can be the integrated sum of the GW SED. While the early-expansion model (inflation and phase transitions) is fundamental to the PGW amplitude evolution, the late stage has also an important impact on the low-frequency GW (Bhoonah et al. 2020). Recently found NANOGrav 12.5 yr results might reflect some of the PGWs as discussed in Bian and Zhou (2020).

(2) x particles and dark matter

As mentioned above, the x particles can be the best candidate of dark matter. It will be heavier than quarks, and will almost not be interacting with normal matter. However, the same as normal matter, the x particles have masses and gravitational effects. As we assumed, if the x particles are bosons, they can form Bose-Einstein condensed (BEC) matter. It is also possible the BEC matter can form black holes (x-BHs) in the universe before the formation of first stars, i.e. the primordial BHs (PBHs), e.g. Cai et al. 2020, De Luca et al. (2020).

In the standard cosmology model, the dark mater occupies ~23% of the energy of universe, much more than normal matter, so they dominates the large scale structure formation of the universe, e.g. galaxies and galaxy clusters. If the x particles are the dark matter, they will be distributed not only in galaxies but mainly in the galactic halos and galaxy clusters. Because of weak or no interacting with normal matter through the strong, weak, and electromagnetic interactions, we could not observe them with the electromagnetic waves or neutrinos. They can be observed by the gravitational effects, e.g. the galactic rotational curve, which is one of key evidences of dark matter.

The x particles can be in the forms of isolated particles, x-BHs, and the Bose-Einstein condensed matter. It is not clear which form is the main component dominating the dark matter. From micro-lensing observations from galactic halos, it seems less possible that the x-BHs are the dominant component of dark matter. The x particles and some condensed x-matter could be the main components of dark matter.

Except the galactic gravitational curves, the light from the annihilation of x particle and its anti-particle would be observable. Assuming the lower limit of mass of x particle is a few times of quarks, e.g., similar to the mass of neutron, i.e. nearly 1 GeV, the signal wave of the annihilation of x particles will be > 1 GeV. So, to find such gamma ray signals is important for the x particles. This reaction may need a very high speed of the x particles for collision. But there is a suggestion that the dark matter particle has an energy of > 300 GeV (Chan et al. 2019). Another way is to find the gravitational accretion of normal matter onto the BEC matter or x-BH in galactic halos, the accreting normal matter can emit electromagnetic waves that may be observable.

The x particles may be also condensed in galactic centers and star clusters, which can also help to form the structures. It is reported that there is a GeV signal from the Galactic center (Kosack et al. 2004), might be a signal of x particles annihilation. There is also a gamma-ray background in the universe, some of them might come from the x particle annihilation. However, if the mass of x particle is much heavier than neutron, the gamma ray from the x particle annihilation will go up to TeV as suggested in Chan et al. (2019), which will need very high-energy telescope to observe.

Indirect observations, like the scattering of quasar light by the x particles may be possible, and also quasar light can be lensed by the BEC matter to cause the micro-lensing effects. So, one could be able to detect the microlensing effects in our Galaxy.

The x particles are assumed to be created in the abrupt phase-transition of vacuum in space-time earlier than the production of quarks, the latter is produced by subsequent phase-transitions of vacuum. The number of x particles could be assumed in the form of $N_x = N_0 t^{-m}$, where the number will be almost not changed (m=0) when the universe becomes transparent (after 3 minutes from Big-bang). Before that time, the number of x particles may be declined (m>0) due to frequent annihilation, which might have contributed to a high frequency cosmic background emission which is different from the traditional CMB.

3. Discussion and Summary

In the early phase of the Big-bang universe, the vacuum fluctuations of space-time and the vacuum symmetry breaking (or phase transition) are the main drivers of the production of the primary GWs, the x particles and the normal/standard particles. The products are depending on the scale and energy density of the inflating universe, however, the exact time when the vacuum symmetry breaking (e.g. x particle created) is not clearly known. The vacuum fluctuations and transitions are assumed to be responsible for the dark energy which is taken by the primary GWs.

In fact, the early universe does not collapse into a BH, but explodes as a 'white hole', this means that the inflating universe might be not isotropic, which may have led to the symmetry breaking or the abrupt phase-transition of vacuum, and the vacuum fluctuations.

It is not sure that if the annihilation of x particle is somewhat related to a weak or strong interaction or not, if yes, its relic particles and emission would be observed as signatures of dark matter.

In summary, the observational evidence points to the origin of our Universe from a Big-bang, the normal matter we observed can be well explained by the particles created in the strong and weak interacting phases of the early universe. The dark energy and dark matter, which occupy major contents of the universe, may be created from earlier times than the creation of quarks. We propose that the dark energy is an integration of the GWs in early universe, and the dark matter is composed of x particles which are much heavier than quarks. The x particles may be heavy bosons with little interaction with normal matter except gravity, and they can be free particles or BEC matter as the dominant matter in galaxy clusters and galactic halos. I am new in this field, this note is a framework of ideas for my future work in the field.

Acknowledgement
This work is supported by the National Key R&D Program of China under grant number
2018YFA0404602, and the Key Laboratory of Radio Astronomy, Chinese Academy of Sciences.